\documentclass{article}
\begin{document}
\title{Calculable membrane theory}
\author{C.Lovelace\\ Physics Dept., Rutgers University,\\
        Piscataway, NJ 08854-8019.\\ [9pt]
        lovelace @ physics.rutgers.edu }
\date{February 13, 2009}
\maketitle
\begin{abstract}
The key to membrane theory is to enlarge the
diffeomorphism group until 4D gravity becomes almost
topological. Just one ghost survives and its central 
charges can cancel against matter. A simple bosonic
membrane emerges, but its flat $D = 28$ target space 
is unstable. Adding supersymmetry ought to give
calculable (2,2) membranes in 12 target dimensions,
but (2,1) membranes won't work.
\end{abstract}

\section{Introduction}                          

This paper claims to construct a \emph{calculable} 
membrane theory, unlike too many previous proposals.
My prototype has no fermions and an unstable vacuum, 
but realistic membrane theories should soon follow. 

Let $(s,t)$ be the number of (space,time) dimensions.
We all agree that string theory in (9,1) must be part
of a larger membrane theory, which nobody has yet been
able to construct. Most believe in M theory where
(2,1) membranes wiggle in (10,1) spacetime. A few
heretics, willing to contemplate two times, favor
F theory \cite{vafa} in (10,2). A purely bosonic (2,2)
membrane should then inhabit (26,2). One natural
choice for its internal field theory was conformal
gravity \cite{fradkin}. Schmidhuber \cite{schmid}
tried that with catastrophic results. The critical
dimension was $D = -1538$ \cite{anton} instead of
the desired $D = 28$. An alternative is topological
4D gravity (many attempts reviewed in \cite{constant}),
where tables of proliferating ghosts cover the page
like armies of ants, all engaged in Mutually Assured
Destruction. Not surprisingly topological gravity
itself died: \cite{constant} has no citations and
there are few subsequent papers. Resurrection only 
became feasible when I discovered contraception for ghosts.
My construction of the bosonic membrane will use (4,0)
signature. However Majorana-Weyl fermions require
$s=t\!\pmod{8}$ .

Section 2 is didactic. I show how topological
Yang-Mills (YM) theory can be reformulated with no
secondary ghosts. This yields nothing new, but
illustrates my procedure on a familiar example.
Section 3  then introduces the new ideas.

In section 4  I analyse topological 4D gravity.
In this case a new theory results. One can consistently
topologize nine components of the metric tensor, while
quantizing the conformal mode only.

In section 5  I identify the latter with the Liouville 
field previously used to compensate conformal anomalies 
of matter. It is actually a fourth order ghost coupled to
Q-curvature. It plays the same role in membrane theory
that the bosonized ghost plays in string theory. 
Adding 28 free scalars gives a membrane equivalent
of the bosonic string, equally calculable but
with similar disabilities.

Supersymmetry is postponed to a future paper, but
section 6 sketches how a complete membrane theory
would work.

\section{Eliminating secondary ghosts}             

First I need some notation. In Euclidean 4D space
the Lorentz group is $SU(2) \times SU(2)$. Its
irreducible representations $<\!m,n\!>$ are 
labelled by two integers (twice the spins). The
representation has dimension $(m+1)(n+1)$. Thus
$<\!1,1\!>$ is a vector. An antisymmetric tensor
is $<\!2,0\!>\,+\,<\!0,2\!>$ = selfdual(SD) +
antiselfdual(ASD). A field carries $m$ totally
symmetric left spinor indices $AB\ldots$ 
each with 2 values, and $n$ totally symmetric 
right ones $A'B'\ldots$ . Thus a vector index
decomposes $\mu \;=\; AA'$. Spinor indices are
raised or lowered with $\epsilon\;=\; 
{ 0 \quad\; 1 \choose -1 \;\; 0 } $.
Further details can be found in \cite{penrose}.
I will use $\to$ for differential operators,
$\mapsto$ for conformal transformations.

Write the Yang-Mills(YM) potential as a matrix 
in the adjoint representation of the gauge group
\begin{equation}\label{e1}                            
    \hat{\Phi}_\mu^{de}\;\equiv\;gA_\mu^a\,f^{dae}\;,
\end{equation}
where $f^{dae}$ are structure constants and $g$ is
the coupling. Define
\begin{equation}\label{e2}                            
    \hat{\nabla}_\mu\;\equiv\;\hat{1}\:\partial_\mu
                     \;+\; \hat{\Phi}_\mu \;.
\end{equation}
Then
\begin{equation}\label{e3}                            
    [ \hat{\nabla}_\mu \;,\; \hat{\nabla}_\nu ]
    \; =\; \hat{F}_{\mu\nu} \;.
\end{equation}
The usual $\Lambda$ gauge is
\begin{equation}\label{e4}                            
    \delta \hat{\Phi}_\mu  \;=\;
    [ \hat{\nabla}_\mu \;,\; \delta\hat{\Lambda} ] \;.
\end{equation}

We would like to make this theory topological.
Traditionally \cite{topo} one introduces a shift
symmetry $\delta\Phi_\mu\; = \mbox{all ghost }$,
and then uses secondary ghosts to disentangle 
the gauge $\delta\Lambda$. This works for YM,
but leads to extreme complexity in gravity.

To find an alternative, return to fundamentals.
Topological field theories grew out of instanton
physics. An (anti)instanton is a solution of the
classical equation $F(SD) = 0$. It depends on a
finite number of parameters. These can be counted
by introducing an elliptic complex \cite{eguchi},
p.322. This consists of two differential operators
$D_0\;=\;\delta\Phi/\delta\Lambda$ and
 $D_1\;=\;\delta F(SD)/\delta\Phi$ . Explicitly $D_0$ 
is (\ref{e4}), and $D_1$ in bispinor notation is
\begin{equation}\label{e7}                            
    \delta\hat{F}_{AB} \;=\; 
                  [ \hat{\nabla}^{C'}_{(A} \;,
                    \;\delta \hat{\Phi}_{B)C'} ] \;.
\end{equation}
where (AB) means symmetrize. The operators form a
complex $D_1.D_0\;=\;0$ . This just says
that the instanton equation $F(SD) = 0$ is gauge
invariant. The complex is elliptic because
$D_1^*.D_1\;+\;D_0.D_0^*$ is an elliptic differential
operator. Here * means adjoint. The index of this
elliptic complex then counts the number of instanton
parameters. In terms of Lorentz representations,
the complex can be written
\[
    <\!0,0\!>\;\to\;<\!1,1\!>\;\to\;<\!2,0\!> .
\]
The adjoint operators reverse the arrows
\[
    <\!2,0\!>\;\to\;<\!1,1\!>\;\to\;<\!0,0\!> .
\]
Using cyclic symmetry of the trace and spinor
index rules, one finds that $D_1^*$ is
\begin{equation}\label{e5}                            
    \delta \hat{\Phi}_{AA'}  \;=\;
    [\hat{\nabla}^B_{A'}\;,\;\delta\hat{J}_{(AB)} ] \;.
\end{equation}
and $D_0^*$ is
\begin{equation}\label{e6}                            
    \delta\hat{L}\;=\;-[\hat{\nabla}^\mu\;,
                    \;\delta\hat{\Phi}_\mu]\;,
\end{equation}

Now I can present my new method of making
a field theory topological. $J_{AB}$ in (\ref{e5})
is a selfdual Hertz potential with 3 components.
Use it as an \emph{additional} gauge invariance.
Because $D_1.D_0\;=\;0$ , it is orthogonal to the
old $\Lambda$ gauge (\ref{e4}). We now have enough
gauges to cancel all four components of $\Phi_\mu $ .
Use $D_1$ to fix the $J$ gauges by $F(SD) = 0$ , and
$D_0^*$ to fix the $\Lambda$ gauge by $L = 0$.
Next employ standard Faddeev-Popov quantization.
We have four ghosts $C_{AB}\;+\;C_0$ (where 0 means
no index), four antighosts $B^{AA'}$, four Lagrange
multipliers $N^{AB}\;+\;N^0$, and four fields
$\Phi_{AA'}$. The Lagrangian is
\begin{equation}\label{a1}                     
     \hat{B} (D_0 + D_1^*) \hat{C}\;
      +\;\hat{N} (D_0^* + D_1) \delta\hat{\Phi}\; ,
\end{equation}
traced over the internal gauge group.
The first term is fermionic, the second bosonic, and
they contain $4\times 4$ Lorentz matrices that are
\emph{adjoints}. In (2,2) signature the Lorentz group is
$SL(2,R)\times SL(2,R)$, so everything is real,
the adjoints become transposes and the determinants
obviously cancel as they should for a topological
field theory. Note that I have written $\delta\hat{\Phi}$
not $\hat{\Phi}$. The Lagrangian (\ref{a1}) is actually
in tangent space. This will be important for gravity.
People more mathematical than I see a relation to
balanced topological theories \cite{dijk} .

\section{Main ideas}                              

In this section I outline my new theory. In the next
two sections I will fill in the mathematical details.

Suppose we instead perturb a gravitational instanton.
The diffeomorphism generators $\delta\xi^\mu$ are $<\!1,1\!>$ .
In tangent space we can decompose $\delta g_{\mu\nu} $
into $<\!0,0\!> + <\!2,2\!>$ = trace + traceless. The 
curvature $\delta R_{\mu\nu\rho\sigma}$
decomposes \cite{penrose} into $<\!0,0\!> + <\!2,2\!> +
<\!4,0\!> + <\!0,4\!>$. The first pair make up the Ricci tensor,
the second pair the Weyl tensor which splits SD + ASD.
The gravitational instanton complex is
\begin{equation}\label{a3}                        
    \mbox{diffeos}\;\to\;\mbox{metric}\;\to\;
    \mbox{SD Weyl} \; \to \; 0\;.
\end{equation}
Thus $D_0$ maps $<\!1,1\!>\:\to\:<\!2,2\!> + <\!0,0\!>$, 
$D_1$ maps $<\!2,2\!>\:\to\:<\!4,0\!>$. Now use the
adjoint operators to introduce new multiplets as we did
for YM. First we have a 5 component $<\!4,0\!> \:J$ 
multiplet mapped into the metric tensor by $D_1^*$ : 
$<\!4,0\!>\:\to\:<\!2,2\!>$. As with a Kaehler potential, 
$g$ contains its second derivatives. Use $J$ as a gauge 
to enlarge the diffeomorphism group. Fix these $J$
gauges with the 5 component $W(SD) = 0$. Finally use
$D_0^*$ : $<\!2,2\!>\:\to\:<\!1,1\!>$ to fix the
diffeomorphisms with 4 harmonic coordinate conditions
like (\ref{e6}). 

Equation (\ref{a1}) is still consistent 
because both matrices are now $9 \times 9$, but 
four $<\!1,1\!>$ diffeomorphisms plus five $<\!4,0\!>\, J$ 
gauges are not enough to cancel the ten components of
$\delta g_{\mu\nu}$. Its $<\!0,0\!>$ trace mode survives.
This was not easily seen before, because the secondary 
ghosts were so complicated. The 4 diffeo ghosts 
\emph{alone} could not have been cancelled \cite{anton} 
because $9 + 1$ is the only covariant way to split 
$\delta g_{\mu\nu}$ .

Next consider conformal transformations $g_{\mu\nu}
\mapsto e^{2\lambda} g_{\mu\nu} $, denoted by
$\mapsto$. If $\delta\lambda$ is
infinitesimal, it will only change the $<\!0,0\!>$ trace
component of $\delta g_{\mu\nu}$, not the $<\!2,2\!>$
part that we fixed. The situation is now analogous
to 2D, where only the conformal mode survives after
diffeomorphisms have been fixed. The analogy can be
pushed further. In any even dimension Branson
\cite{branson} constructed a scalar function $Q(R)$
of the curvatures, and a linear differential operator
$P$. Under a conformal transformation, $Q \mapsto Q +
P\lambda$. This makes it very natural to introduce
a scalar field $\sigma(x)$ with Lagrangian
$\frac{1}{2} \sigma P \sigma + Q \sigma$. Then $\sigma
\mapsto \sigma + \lambda$ and has propagator
$\sim \log x$ allowing it to be exponentiated.
In two dimensions $Q = R$, the scalar curvature,
and $P$ is the Laplacian. String theorists should
immediately recognize $\sigma$ as the bosonized ghost
\cite{fms}, whose $c = -26$ conformal anomaly fixes
the critical target dimension. Its 4D analog has just
the right properties to serve as the missing 
$<\!0,0\!>$ component of $\delta g_{\mu\nu}$.

In four dimensions $P = \Delta_4$ was independently
discovered by Riegert \cite{riegert}. He used it
to write the most general curved space anomaly
$\Gamma(R)$ as a local action $\int [\frac{1}{2}
\sigma \Delta_4 \sigma + \Gamma(R) \sigma] $.
He mistook $\sigma$ for a Liouville field, but a
fourth order scalar can only be a ghost. Then
$\sigma$ contributes its own conformal anomaly
with $a = -28$. This is titillating, but was thought
to be useless because the diffeomorphism ghosts
would add $a = 1566$ \cite{anton}. My version of
topological gravity kills the old diffeo ghosts
and the new $J$ ghosts, but not the $\sigma$ field,
so the two pieces fit together perfectly. This
solves another puzzle \cite{berk}: twistor strings
couple to conformal gravity with $c = 28$ matter,
which is impossible in standard conformal gravity
\cite{fradkin}.

In this paper I will just consider the simplest
bosonic membrane in $D = 28$, but it seems very
likely that adding supersymmetry will construct
F theory in $D = 12$, and it will be CALCULABLE.

\section{Topological gravity reexamined}           

Now look at some equations. I will use semicolons for
covariant derivatives and \cite{weinberg,penrose,riegert}
curvature sign, opposite to \cite{anton,perry,carneiro}. 
In \emph{tangent space} Lorentz decomposition
is covariant, eg. $g^{\mu\nu} \delta g_{\mu\nu} $
is the $<\!0,0\!>$ part of $\delta g $.

Our first differential operator
$D_0$ maps infinitesimal diffeomorphisms
$\delta\xi^\mu (x) $ into the traceless part of
$\delta g_{\mu\nu} $. It is well known
\begin{equation}\label{b1}                         
    \delta g_{\mu\nu} = D_0.\delta\xi \; =\;
    \delta\xi_{\mu ; \nu} + \delta\xi_{\nu ; \mu}
    - \frac{1}{2} g_{\mu\nu}
                 \delta\xi^\alpha_{; \alpha} \;.
\end{equation}
Our second operator is
\begin{equation}\label{b2}                        
    D_1\; =\; \delta W(SD) / \delta g_{\mu\nu}\;,
\end{equation}
where $W(SD)$ is the selfdual half of the Weyl tensor.
This takes a page to calculate. The Palatini identity
\cite{weinberg},p.290 can be generalized to
\begin{equation}\label{b3}                        
    2 \delta R_{\lambda\mu\nu\kappa} =
      \delta g_{\lambda\nu ; \mu\kappa}
    + \delta g_{\mu\kappa  ; \lambda\nu}
    - \delta g_{\lambda\kappa ; \mu\nu}
    - \delta g_{\mu\nu     ; \lambda\kappa}
    + R^\alpha_{\;\mu\nu\kappa} \delta g_{\lambda\alpha}
    - R^\alpha_{\;\lambda\nu\kappa} \delta g_{\mu\alpha} .
\end{equation}
In bispinor notation with $a = AA'$, the curvature
decomposes \cite{penrose}
\[                                                
    R_{abcd}\;=\;X_{ABCD}\;\,\epsilon_{A'B'}\;\,\epsilon_{C'D'}
        \;+\;\Phi_{ABC'D'}\;\,\epsilon_{A'B'}\;\,\epsilon_{CD} 
\]
\begin{equation}\label{b4}
    +\;\Phi_{CDA'B'}\;\,\epsilon_{AB}\;\,\epsilon_{C'D'}\;+\;
    \widetilde{X}_{A'B'C'D'}\;\,\epsilon_{AB}\;\,\epsilon_{CD}\;,
\end{equation}
and $X$ further splits
\begin{equation}\label{b5}                       
    X_{ABCD} = W_{ABCD} \;+\; \frac{1}{24} R\;
               (\epsilon_{AC}\;\,\epsilon_{BD}\;+\;
                \epsilon_{AD}\;\,\epsilon_{BC} ) \;.
\end{equation}
There is a similar split of $\widetilde{X}$ into
a different $\widetilde{W}_{A'B'C'D'}$
but the same $R$. Here
\begin{equation}\label{b6}                       
    \epsilon_{AB} \;=\; \epsilon^{AB} \;=\;
             { 0 \quad\; 1 \choose -1 \;\; 0 }
\end{equation}
raises and lowers spinor indices. $W$ and 
$\widetilde{W}$ are the SD and ASD halves
of the Weyl tensor, $\Phi$ is the traceless
part of the Ricci tensor, and $R$ is the scalar
curvature. To project $\delta W_{ABCD}$ 
from (\ref{b3}), substitute (\ref{b4}) 
and contract with
\begin{equation}\label{b7}                      
    \frac{1}{4} (\epsilon^{A'C'}\;\epsilon^{B'D'}
              \;-\;\epsilon^{A'D'}\;\epsilon^{B'C'})\;.
\end{equation}
This gives $\delta X_{ABCD}$, and symmetrizing
then gives $\delta W_{ABCD}$. Rewriting\\ $;a \equiv
\nabla_{AA'}$ for covariant derivatives,
$D_1$ is
\begin{equation}\label{b8}                     
    \delta W_{ABCD}\; = \;\frac{1}{2}\;
    \nabla_{(A}^{A'} \nabla_B^{B'}\,
                     \delta g_{CD)A'B'} +
    \frac{1}{2}\; \Phi^{A'B'}_{(AB}\,
                     \delta g_{CD)A'B'} \;,
\end{equation}
where $(ABCD)$ means symmetrize. This was 
written down before \cite{perry,witten,itoh}
and could also have been deduced from the Bach
tensor (\cite{penrose} v.2, p.127). $D_0$ and
$D_1$ have been rigorously proved to form an
elliptic complex \cite{itoh,king}. Its index
(\ref{f2}) was calculated long ago 
by Singer (quoted in \cite{eguchi}, p.369) .

The adjoint $D_1^*$ is easily found \cite{perry}
\begin{equation}\label{b9}                        
    \delta g_{ab}\;=\;(D_1^*.\delta J)_{AA'BB'}  =
    \frac{1}{2} [\nabla^C_{(A'} \nabla^D_{B')}
    + \Phi^{CD}_{A'B'} ]\; \delta J_{(ABCD)} \;.
\end{equation}
This equation is my gold nugget from topological gravity.
$J_{ABCD}$ is totally symmetric in its four left spinor
indices. Thus its five components are the new gauges
orthogonal to $\delta\xi_{AA'}$ in (\ref{b1}) that make
gravity almost topological. Standard Faddeev-Popov
quantization now gives 9 ghosts $C_{AA'} + C_{ABCD}$,
9 antighosts $B^{AA'BB'}$, and 9 Lagrange multipliers
$N^{ABCD} + N^{AA'}$. The latter impose
\begin{equation}\label{b10}                      
    \delta W_{ABCD}\; =\; 0 \;,
\end{equation}
and
\begin{equation}\label{b11}                      
    0\;=\;(D_0^*. g)_{AA'}\;=\;-2\:\delta g^b_{a;b}
      + \frac{1}{2} \:\delta g^b_{b;a} \;,
\end{equation}
which is just the harmonic coordinate condition with
its trace removed. Note that the gauge fixing refers
to tangent space ($\delta g$), and must now be 
integrated along some deformation path from an
instanton solution. If we start from
flat space, we can just remove $\delta$ and replace
semicolons by commas in (\ref{b10})-(\ref{b11}), but
there are other possibilities.

In split (2,2) signature the Lorentz group is
$SL(2,R) \times SL(2,R)$ , so everything is real, the
adjoints become transposes and the $9 \times 9$ 
determinants in the partition function of
\begin{equation}\label{b12}                      
    \mathcal{L}\; =\; B (D_0 + D_1^*) C
               \; +\; N (D_0^* + D_1) \delta g
\end{equation}
must cancel. In Euclidean signature there might be
phase factors remaining.

However the $<\!0,0\!>$ trace component
\begin{equation}\label{b13}                     
    \delta g_{\mu\nu}\; =\; g_{\mu\nu}\:
    [\delta\xi^\alpha_{;\alpha}/2\:+\:2\delta\lambda ]
\end{equation}
has clearly not been cancelled. The first term is the
subtraction in (\ref{b1}), which doesn't change the
curvatures, while $\delta\lambda$ is an extra
conformal transformation which does change the
Ricci tensor.

\section{The conformal mode}                          

Next recall some string theory. In two dimensions all
surfaces are conformally flat $g_{\mu\nu} = e^{2\lambda}
\delta_{\mu\nu}$ , and the Einstein action is a total
derivative
\begin{equation}\label{c1}                            
    \sqrt{g} R\; =\; 2 \lambda_{,\alpha\alpha} \;.
\end{equation}
Its integral is the Euler number
\begin{equation}\label{c2}                            
    \chi = - \frac{1}{4\pi} \int d^2 x \sqrt{g} R
       \;=\; 2(1 - \gamma) \;,
\end{equation}
which depends only on the genus $\gamma$. A conformal
transformation can move curvature around and even
concentrate it at singular points, but it cannot change
the total. The wave equation of the bosonized ghost is
\cite{fms}
\begin{equation}\label{c3}                            
    \sigma_{,\alpha\alpha}\;=\; \frac{3}{4}\sqrt{g}R
    \;=\; \frac{3}{2}\:\lambda_{,\alpha\alpha} \;,
\end{equation}
so under a conformal transformation it acquires a new
classical term $\sigma \mapsto \sigma + \frac{3}{2} \lambda$,
which by (\ref{c1}) looks like an electrostatic potential
with curvature as charge density. This is the background
charge phenomenon \cite{fms}. The factor $3/2$ in
(\ref{c3}) is related to its conformal anomaly
\begin{equation}\label{c4}                            
    c \;=\; -26\; =\; 1 - 12 \:(3/2)^2 \;,
\end{equation}
which determines the critical dimension of the matter 
fields. The number of moduli of a Riemann surface is
(\cite{polch} v.1, p.152)
\begin{equation}\label{f1}                            
    \mu \;=\; \kappa - 3 \chi \; ,
\end{equation}
where $\kappa$ is the number of conformal Killing
vectors. A sphere has $\mu = 0, \kappa = 6, \chi = 2$,
which determines its total background charge to be 
-6, and leads to the 3/2 in (\ref{c3})-(\ref{c4}).

Traditionally the $\sigma$ ghost is a bosonized
version of the bc ghosts. This is a gauge-dependent
statement. In 2D, $\delta\xi$ and the traceless part
of $\delta g$ both have 2 components. Therefore in
traceless harmonic gauge $D_0^*.\delta g = 0$, the bc
ghost determinant cancels against its gauge-fixing
term, like (\ref{b12}) with $D_1$ omitted. However
the cancellation is incomplete -- a finite number
of zero modes remain, which are described by the
moduli formula. The 2D $\sigma$ Lagrangian then
bosonizes these zero modes only. It could have been
inferred entirely from Q-curvature and the constant
in the moduli formula, which is how one must proceed
in 4D.

In 2D the only possible gravitational trace anomaly
is (\ref{c1}), but in 4D there are four scalars
containing four derivatives of $g$: 
$\nabla^2 R$ , $R^2$,
\begin{equation}\label{c6}                            
    F\; \equiv\; [W(SD)]^2 + [\widetilde{W}(ASD)]^2 \;,
\end{equation}
which is the square of the Weyl tensor from (\ref{b4})
-(\ref{b5}), and
\begin{equation}\label{c7}                            
    E\; \equiv\; R^2_{\mu\nu\rho\sigma}
                 - 4 R^2_{\mu\nu} + R^2 \;,
\end{equation}
which integrates to the Euler character
\begin{equation}\label{c8}                            
    32 \pi^2\; \chi = \int d^4 x \sqrt{g}\: E \;.
\end{equation}
A second pseudoscalar total derivative gives
the Hirzebruch signature
\begin{equation}\label{c9}                            
    48 \pi^2\; \tau = \int d^4 x \sqrt{g}\:
       \{ [W(SD)]^2 - [\widetilde{W}(ASD)]^2 \} \;.
\end{equation}
Both have extra boundary terms \cite{eguchi}.
Note that $F$ becomes topological if the Weyl 
tensor is either SD or ASD.

Riegert \cite{riegert} investigated cures for these
anomalies. The first $\nabla^2 R$ is easily cured
because it can be derived from the action $\int R^2$.
The second $R^2$ is incurable and would mean the
theory was inconsistent. The interesting cases are
$F$ and $E$, which can be cured by introducing a new
scalar field. To see why, examine their conformal
transformation properties (\cite{carneiro} with $R\to -R$).
Define the fourth order Paneitz-Riegert operator
\begin{equation}\label{c10}                          
    \Delta_4\;\equiv\;(\nabla)^4 - 2 R^{\mu\nu}
    \nabla_\mu \nabla_\nu  +  \frac{2}{3} R\:(\nabla)^2
    -  \frac{1}{3} R_{;\mu} \nabla^\mu  \;,
\end{equation}
and Q-curvature \cite{branson}
\begin{equation}\label{c12}                          
    Q\;\equiv\;\frac{1}{4}\,(E+\frac{2}{3}\nabla^2 R)\;.
\end{equation}
Then under a conformal transformation
$g_{\mu\nu} = e^{2\lambda} \hat{g}_{\mu\nu}$ ,
$\sqrt{g}\Delta_4$ and $\sqrt{g}F$ are invariant,
while \cite{carneiro}
\begin{equation}\label{c13}                          
    g^{1/2}\: Q\; =\; \hat{g}^{1/2} \,
    (\hat{Q}\; +\; \hat{\Delta}_4 \lambda )\;.
\end{equation}
With the gauge fixing conditions (\ref{b10})-(\ref{b11})
we can separate the unfixed trace part $\sigma$
of the metric tensor
\begin{equation}\label{c14}                          
    g_{\mu\nu}\;=\;e^{2\sigma(x)}\;\hat{g}_{\mu\nu}
\end{equation}
from the fixed part $\hat{g}$ . Then \emph{classically}
the $\sigma$ field is just a conformal transformation
$\lambda$. Its kinetic term is $\sigma \Delta_4 \sigma$
and its source \emph{might be} the conformal anomaly
\begin{equation}\label{c5}                            
    16 \pi^2\sqrt{g} < T^\alpha_\alpha > \;\equiv\;
    \delta\Gamma_{ren}(g) / \delta\lambda \; =\;
    \sqrt{g} \:( cF\: -\:4 aQ ) \;,
\end{equation}
where $\Gamma_{ren}(g)$ is the renormalized vacuum 
amplitude.
The coefficients $c$ and $a$ in (\ref{c5}) are two
central charges, which depend on the matter theory,
and are known in many cases \cite{anselmi,asorey}.
Riegert \cite{riegert} and his followers \cite{anton}
used $\sigma$ as a Liouville field, including the
central charges of the matter fields in its
Lagrangian by (\ref{c5}), and \emph{then} quantizing. 

However this is not how critical strings
work. There matter and ghosts are completely
independent and their central charges cancel linearly.
So in analogy to (\ref{c3}), whose source term is not 
a central charge of anything, postulate a separate 
$\sigma$ Lagrangian to describe zero modes left over
from the cancellation (\ref{b12}).
\begin{equation}\label{c15}                          
    16 \pi^2\; \mathcal{L}_\sigma \; = \;
    \alpha\:\sigma \hat{\Delta}_4 \sigma \;+\;
    \beta\:\hat{Q} \:\sigma \;+\; \gamma\:\hat{F} 
    \:\sigma \; .
\end{equation}
Then $\alpha = -1/2$ gives $\sigma$ the propagator 
$\log (x^2)$. This is essential because it will be
exponentiated. The number of moduli of a
gravitational instanton is \cite{king,itoh,perry}
\begin{equation}\label{f2}                            
    \mu \;=\; \kappa \;+\; (29 |\tau| - 15 \chi)/2 \;.
\end{equation}
This is the exact analog of (\ref{f1}). The sphere
again has $\mu = 0$ because $\kappa = 15$ cancels 
$\chi = 2$ . The constants $\beta$ and $\gamma$ in
(\ref{c15}) should therefore be related to the numbers
15 and 29 in (\ref{f2}). This will give the correct
background charge for various instantons. Probably
$\beta = 15/4$ .

At this point we encounter a sharp difference from
2D, where $\beta = 3/2$ determines the critical
dimension by (\ref{c4}). Elizalde et al.\cite{eliz} 
calculated the one loop 4D anomaly for a very general 
fourth order $\sigma$ Lagrangian. They found that the 
constants $\alpha, \beta, \gamma$ in (\ref{c15}) had 
no effect on its central charges, which were determined 
entirely by $\Delta_4$ unless the other terms were
nonlinear in $\sigma$. 

A first guess for the membrane matter field is that
the target coordinates are described by free scalars, 
as for strings in flat space. For one fourth order 
ghost $\sigma$ and $D$ free scalars $\phi$, the
central charges are \cite{anton}
\begin{equation}\label{c16}                         
    a = (D - 28)/360 \;, \qquad c = (D - 8)/120 \;.
\end{equation}
The contributions of unitary matter are always positive 
\cite{anselmi} , while those of fourth order scalars 
or third order spinors are negative \cite{asorey} .
We expect from string theory that the renormalization 
group equations of the 4D membrane theory will become 
the classical field equations of target space gravity. 
Then $D \ge 28$ since $a$ never increases \cite{anselmi}; 
and $c \ne 0 $ makes flat 28D space unstable. The 
bosonic string sits precariously on top of the hill,
but the membrane is halfway down and sliding. 
It might flow under a $\phi_4^4$ interaction, but this
is unlikely to be the answer since there is no fixed
point and no geometric interpretation. The correct 4D
matter theory is probably supersymmetric gauge with
the group manifold as its target space.

Though $\sigma$ remains a free quantum field, its classical
background will change $\sigma\;\mapsto\;\sigma + \beta
\lambda$ under a conformal transformation, just like
the 2D bosonized ghost, so we will again have background
charge coming from Q-curvature (\ref{c7})-(\ref{c12}).

\section{Prospects}                                  

Reexamination of 4D topological gravity showed that
it was not quite topological. Nine components of
$g_{\mu\nu}$ were fixed, but the tenth survived
to become a conformal mode $\sigma$. This was coupled
to Q-curvature and used to cancel matter anomalies,
just like the 2D bosonized ghost. We know from 2D that
supersymmetry will add another ghost \cite{fms}
with more possibilities for anomaly cancellation.

The gauge fixing conditions (\ref{b10})-(\ref{b11})
split the theory cleanly into a topological sector
involving the traceless part of $\delta g_{\mu\nu} $,
and a physics sector involving $\sigma$ and its
cancelling matter fields.. Doing BRST in the
topological sector would be a pointless duplication
of rigorous mathematics \cite{king,itoh}. Only
the physics sector needs a nilpotent BRST operator.
Analogy to the 2D bosonized ghost, \cite{peskin}
eq.(4.19), suggests (* = dual 3-form)
\begin{equation}\label{d1}                           
    BRST\;=\;\oint e^{\alpha\sigma}
              x^\mu T_{\mu\nu} (dx^\nu )^* \;.
\end{equation}
Here $ T_{\mu\nu} $ is the combined energy-momentum
tensor of the matter fields and $\sigma$. The
$e^{\alpha\sigma}$ factor is allowed because of its
$\log (x^2)$ propagator. The constant $\alpha$
will be determined dimensionally.  Nilpotence could 
be checked by short distance expansion in the tangent 
space, and would probably be equivalent to vanishing 
of all central charges. A scalar ghost with $\log x$ 
propagator only exists in even dimensions. None of 
this would work for the 3D membrane of M theory.

The next step is to couple the membrane to external
sources. For a Euclidean worldsheet the asymptotic
particles sit on punctures. Very natural 4D equivalents
are nuts and bolts \cite{gibbons}. These are movable
coordinate singularities (like the poles of a sphere)
which contribute to $\chi$ and $\tau$. A pointlike
nut would carry a single particle, a 2D bolt would
carry the entire worldsheet of an asymptotic string.
A bolt is the Euclidean continuation of a black hole
horizon, which makes an appropriate perch for an
external string worldsheet.

Even though 9 components of $g_{\mu\nu}$ have been
topologized, they will still contribute classically.
Gravitational instantons correspond to Riemann
surfaces \cite{torre}, and there should be a
propagator and vertices from which they can be
constructed by membrane field theory with the moduli
(\ref{f2}) emerging automatically. The $\sigma$ 
Lagrangian will be very important in providing
consistent measures, and it should soon become
clear whether the Ansatz (\ref{c15}) is correct.

There is a good chance that a supersymmetric extension 
will resolve all difficulties. I hope to write a further
paper on this. Compared to string theory, there will
be relatively few stable vacua, since we now require 
4D not 2D conformal invariance, eg. the 12D target
geometry might be quaternionic Kaehler compactified 
on $G_2/SO(4)$. I will then propose that the second
time is real, and perhaps detectable.

I thank G.Moore for encouragement, W.Siegel for a
provocative 2007 correspondence, and especially
arXiv from which I downloaded over 2000 papers,
sifting like a goldminer for the secret of
F theory. I am grateful to N.Berkovits, S.Deser, 
E.Diaconescu, G.Moore, A.Schwimmer,
and M.Strassler for comments. A.Schwimmer was
especially helpful in improving the first version.

\end{document}